\documentclass[a4paper,11pt]{article}
\usepackage{pos}

\usepackage{graphicx}
\usepackage{xcolor}

\title{The format for GRAND data storage and related Python interfaces}
\ShortTitle{The format for GRAND data storage and related Python interfaces}

\author*[a]{Lech Wiktor Piotrowski}

\affiliation[a]{Faculty of Physics, University of Warsaw,\\
Pasteura 5, 02-093 Warsaw, Poland}

\onbehalf{for the GRAND Collaboration}

\emailAdd{science@lwp.email}

\abstract{
The vast amounts of data to be collected by the Giant Radio Array for Neutrino Detection (GRAND) and its prototype — GRANDProto300 — require the use of a data format very efficient in terms of i/o speed and compression. At the same time, the data should be easily accessible, without the knowledge of the intricacies of the format, both for bulk processing and for detailed event-by-event analysis and reconstruction. We present the format and the structure prepared for GRAND data, the concept of the data-processing chain, and data-oriented and analysis-oriented interfaces written in Python.
}

\ConferenceLogo{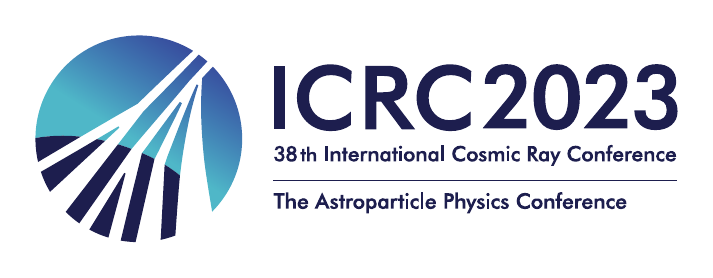}

\FullConference{%
38th International Cosmic Ray Conference (ICRC2023)\\
  26 July - 3 August, 2023\\
  Nagoya, Japan}

\begin{document}

\maketitle

\section{Introduction}
\label{sec:Introduction}

The Giant Radio Array for Neutrino Detection (GRAND) \cite{GRAND:2018iaj} is a planned large-scale radio-observatory of ultra-high-energy (UHE, $E \gtrsim 10^8$ GeV) cosmic particles: cosmic rays, gamma rays, and neutrinos. UHE particles initiate extensive air showers (EAS) in the Earth's atmosphere, whose radio signature is observable using affordable radio antennas. In its final stage, GRAND would reach a huge exposure with 20 arrays of 10,000 radio antennas, each occupying 10,000 km$^2$, adding up to 200,000 km$^2$ gathered in favourable radio-quiet sites worldwide. We expect huge amounts of data to be collected at the final stage of GRAND and at the earlier stages, including the currently prepared GRANDProto300. This prototype array of 300 antennas will, for the purpose of understanding the radio environment and tuning the trigger, collect a lot of background transient events.

The need for fast access to a large amount of collected data, the need for fast storing of data-processing results, and the need for analysis at different stages, from trigger performance to EAS reconstruction, calls for the use of a very efficient data format, both in terms of size and i/o speed, with a well planned and convenient data structure. At the same time, efficiency should be accompanied by the convenience of data analysis. In the following sections, we describe the type of data we acquire in the experiment and the most common processing chains and analyses performed on it. This leads to describing the chosen data structure and the physical data format. Finally, we describe two Python interfaces to our data. The first one is convenient for processing large quantities of data, and the second is dedicated to detailed, event-by-event data analysis, such as reconstruction of EAS parameters. Both interfaces are part of a Python module for GRAND data handling and analysis --- GRANDlib\footnote{https://github.com/grand-mother/grand}.

\section{The input data and main data-processing chains}
\label{sec:input_data}

The data in the GRAND experiment, like in most experiments of this type, can come from two sources: experimental hardware or simulators. These two data sources differ in content, as detailed below.

Hardware data come in the proprietary binary format generated by the electronics of each of the detector units (DUs) containing antennas. The main part is the ADC traces --- counts vs time bin of the analogue to digital converter, encoding voltage of every channel --- three connected to X, Y and Z arms of an antenna, and one floating channel. The traces from several DUs, together with additional information, such as GPS data and electronics settings, are merged into one event by the central DAQ and stored.

Simulated data contain the value of the electric field of the EAS at the position of a simulated antenna vs time. These traces, along with the input parameters of the simulation, such as the primary particle type, energy and direction, and additional simulation-specific information, such as shower development profiles, are stored in simulator-specific format.

The most common goal of the data analysis in a stable experiment is the reconstruction of the EAS parameters, such as $X_{\rm max}$, energy, and direction, from the hardware data. This reconstruction is based on a reconstructed electric field, which comes from converting the ADC counts to voltage, disentangling experimental influences of cables, filter, LNA, etc., and finally, deconvolving the so-called antenna effective length. Such a field is only an estimation of the real one that reached the antenna and makes the analysis results approximate. Therefore, in the case of simulated data, we do not directly use the electric field provided by the simulator. Instead, simulations are degraded with experimental effects. First, the simulated electric field is converted to analogue voltages according to the antenna's effective length. Then, these voltages are convolved with the responses of all hardware elements, and noise is added. Finally, the result is digitised, yielding a signal mimicking a real voltage measurement. This digitised signal is used as input for reconstructing the simulated electric field.

\section{The data structure and physical format}
\label{sec:data_structure}

\subsection{The selected physical data format}

The GRAND Collaboration has chosen CERN ROOT \cite{Brun:1997pa} TFile and TTrees as the physical format for data storage and processing\footnote{The raw data coming directly from hardware and simulations will also be retained, at least for some, not yet specified, time.}. The format is well-known and widely adopted in the high-rnergy physics community, which requires very fast i/o for very large quantities of data. In our case, the choice was mainly based on:

\begin{itemize}
\item convenience of serialising the data of the type gathered in GRAND;
\item fast output to storage and fast random access during analysis;
\item good responsiveness and support from the developers of the format;
\item Python bindings as the built-in part of the data-format framework;
\item long lifetime expectancy of the data-format framework.
\end{itemize}

Leading contenders were FITS~\cite{fits}, HDF5~\cite{hdf5} and Apache Parquet~\cite{apache-parquet}, which did not fill the above requirements to a large enough degree\footnote{This may change in the future for Apache Parquet, but it remains to be seen how well this wide-application format will suit Astroparticle Physics community in the long run.}. The ROOT TTrees are not perfect though. The ``entry'' structure (usually corresponding to data from a single event or an entire experimental run) in a TTree is fixed for the whole TTree, which is a limitation. Still, on the other hand, this helps to avoid attaching random information at random points throughout the data, resulting in very fast random access times. Another issue is the lack of optimisation of the format for the readout of multidimensional structures, such as 3D C++ Standard Template Library (STL) vectors. In GRAND, we use 3D vectors for storing time traces of an event with \textit{DUs count} $\times$ \textit{antenna arm} $\times$ \textit{time bin} dimensions, where the first and last dimensions are not fixed. ROOT TTrees enforce reading out the whole 3D vector at once, which, however, is expected to be the most common readout case in the data analysis. Finally, the ROOT Python bindings to TTrees follow C++ philosophy and are quite unintuitive to Python users (see sec. \ref{sec:doi}).

\subsection{The data structure}

Our data structure is divided along three axes:
\begin{enumerate}
\item The variability of data: parts constant through the whole run such as detector settings, and parts varying with each event, such as traces
\item The type of data: ADC counts, voltage or electric field measurements, whole EAS information, etc.
\item The source of the data: common for both hardware and simulators or simulator only
\end{enumerate}

These axes are reflected in the name of each TTree type. TTrees named with the TRun* pattern (where "*" is an alphanumeric wildcard) contain information that remains constant for the whole run, each entry corresponding to a separate run, while in T*(without ``Run'') each entry corresponds to a separate event. T*ADC, T*RawVoltage, T*Voltage, T*Efield and T*Shower correspond to information about ADC counts, the voltage measured at the antenna foot, the voltage at the antenna arms, electric fields at the antenna arms and EAS parameters, respectively. T*Sim trees contain information produced {\em only} by simulators, while T*(without ``Sim'') are based on information coming from hardware, which to a large extent can also be obtained from simulators (with unavailable fields left empty). 

The full list of TTrees, which may be extended in the future, is as follows: TRun, TRunVoltage, TRunEfieldSim, TRunShowerSim, TRunNoise, TADC, TRawVoltage, TVoltage, TEfield, TShower, TShowerSim. 

These TTree types do not represent all the combinations of the data structure division axes mentioned above since not all the combinations would hold data.

The standard chain of analysis for the hardware data is: hardware binary blob $\rightarrow$ TRun + TADC + TRawVoltage + TRunVoltage + TVoltage $\rightarrow$ TEfield $\rightarrow$ TShower. The raw (at the foot of the antenna) and final (at the antenna arms) voltages can be obtained from ADC counts at the moment of the conversion of the hardware binary blob with very simple operations. Reconstructing electric field traces stored in TEfield is more complex, with several possible methods and would be done at a later stage of analysis with a separate tool. Finally, the EAS parameters reconstruction can also be done with different methods and stored in TShower at the final analysis stage. The hardware binary blob is currently being converted to ROOT format using a GtoT (GRAND-to-Trees) program in C++ for fast i/o of vast numbers of events provided simultaneously. Consecutive steps are expected (but not enforced) to be done in Python because the number of events involved will be much smaller at these stages.

The standard chain of analysis for the simulators data is: raw simulator data $\rightarrow$ TRun + TEfield + TShower + TRunEfieldSim + TRunShowerSim + TShowerSim $\rightarrow$ TVoltage + TRunVoltage (optionally: $\rightarrow$ TRawVoltage + TADC) $\rightarrow$ TEfield $\rightarrow$ TShower. The first step is just simple storage of simulator data in ROOT format. The conversion to voltage requires convolution with the antenna response to get TVoltage, and then deconvolving it (which is not a simple reversal of the convolution due to the addition of random noise) to get the reconstructed TEfield. Finally, the reconstructed EAS parameters based on the reconstructed electric field are stored in TShower. It is important to remember that the TTree names given here are the names of the types of trees, not their physical instances. Thus, in the simulation chain, there will be two TTrees of type TEfield and two TTrees of type TShower, with precisely the same structure but different content: simulated and reconstructed information. Also, in case of additional processing (for example, smoothing of traces), the resulting data will be stored in a different instance of the same type of tree, with proper information about the processing level and procedure in the metadata. TRunEfieldSim and TRunShowerSim hold simulator-only information constant throughout the simulation run, such as refractivity model or cuts on energies of specific particles. In contrast, TShowerSim holds simulator-only information varying with each event, such as energy and density profiles of different particle types in the EAS.

At the moment of writing this article, it has not been decided what will be the physical distribution of the TTrees to TFiles (corresponding to a physical file on storage). We will create at least a separate file for each data processing stage, but the distribution inside a processing stage is still being discussed.

\section{Data-oriented interface}
\label{sec:doi}

The data-oriented interface (DOI) to our data serves the purpose of fast i/o of our data stored in TTrees but in a Pythonic way, not provided by PyROOT (ROOT to Python bindings). ROOT TTrees, written in C++, assume that TTrees read and write to constant places in memory, which are given as pointers with fixed types specified by the user. In Python, pointers, references and data placement in memory are more abstract concepts. PyROOT interface to TTrees had to work around this fact by, for example, using the first value of an array (from the Array or NumPy libraries) as a placeholder for a scalar value. Using arrays to hold scalar values is far from intuitive and convenient. DOI contains the set of Python classes with names mentioned earlier, allowing using the dot operator (tree\_name.variable\_name scheme) to get and set the values inside the TTree, regardless of the variable type. The TTrees are data containers with additional functionality. Therefore, our classes are Python data classes encapsulating the ROOT TTree.

ROOT TTrees can hold only C/C++ types and classes, and in our case, we hold only simple C types, STL vectors and strings. To use Python scalars, lists and arrays with TTrees, we have initially created three wrapper classes. TTreeScalar accepts and outputs a Python scalar variable but internally holds it as a 1-element NumPy array that keeps the constant address in memory used by the ROOT TTree for input and output. StdString accepts and outputs Python string but internally holds it as an STL string. Finally, the StdVectorList accepts Python lists, NumPy arrays and PyROOT bindings to STL vector, internally holds an STL vector of a specific type, and outputs a Python list. The standard use of a Python dataclass, where the variable is initialised to one of the above types, would not work with TTrees and the tree\_name.variable\_name use scheme, as we need to tell Python to copy the assigned value(s) to a specific place in memory. This can be obtained with getters and setters of the class members, which call a specified function when an assignment operator is used. In our case, they put/read values from a specific place in memory.

However, in our case, using getters and setters leads to a very big code base, where most of the code consists of the same functions, with only the variable name and type replaced. To avoid this issue, we explicitly use custom Python descriptors (as dataclass descriptor-typed fields). Descriptors allow for creating objects with custom behaviour during assignments and readouts. The whole wrapper-class functionality can be coded in the descriptor class, which was done for the simple case of a scalar. However, in the case of StdVectorList, we retain the original wrapper class and use it inside the StdVectorListDesc, to increase readability.

The final outcome is Python classes, which can be used with simple tree\_name.variable\_name assignment and readout scheme with Python variable types, coded as dataclasses without getters and setters, that store and read values from ROOT TTrees. The classes are also iterators over events. In addition, DOI contains DataFile and DataDirectory classes that allow for opening and scanning through a single file/directory, automatic analysis of their contents, gathering essential information and opening of the trees.

\section{Analysis-oriented interface}
\label{sec:aoi}

The physical data structure and distribution are of little importance to many researchers not involved in data processing, but focusing on, for example, EAS reconstruction. For their convenience, we have prepared a set of classes forming the analysis-oriented interface (AOI), which by default automatically loads the most relevant data into an Event class from appropriate sources and distributes and writes analysis results stored in this class into appropriate physical trees, in appropriate physical files. In a manner of speaking, AOI is an analysis-friendly interface to DOI.

The main parts of the Event class are the list of antennas, the list of voltage and electric field traces (for each antenna), the parameters of a simulated shower and the parameters of the reconstructed shower. The data are filled if they are available in the source files. The contents of the Event class are very similar to the tree classes described above. However, they are structured differently, including only the crucial information sub-set, extended by some variables calculated on the fly. For example, the Voltage class encapsulates a Timetrace3D containing voltage traces, adding information on whether the trace was triggered.
Similarly, Efield class encapsulates the Timetrace3D instance corresponding to the simulated or reconstructed electric field trace at the antenna location, adding information about the field polarization angle and geomagnetic to charge excess contribution ratio. The core part of the Timetrace3D class is an array containing \textit{value} $\times$ \textit{time bin} for each antenna arm. While often filled with the values coming from hardware (and processing that followed), the vector type is our internal GRANDlib CartesianRepresentation class based on NumPy ndarray. Among other operations, it allows for easy transformations between coordinate systems -- a feature crucial for some analysis steps. The Timetrace3D class also contains some other information, for example, the time position of each bin, which is ``live''-computed from the stored trace start time and time-bin length, as it is not included in the stored data, due to file size constraints.

The Shower class contains a sub-set of the TShower information for a specific event, but, similarly to Timetrace3D, it puts information such as the Xmax position, direction of the EAS origin and position of the EAS core on the ground into CartesianRepresentation classes for ease of transformations. Similarly, the Antenna class contains the positions and tilts from the vertical orientation of each antenna as CartesianRepresentations, while at the same time adding a link to the used AntennaModel, containing the antennas' effective length, not included in the TTrees. The user can iterate through a list of events using the EventList class.

\section{Summary}
\label{sec:summary}

The structure of the data of the main measurements stored in the GRAND experiment reflects the division into common and simulator-only data, type of data (ADC counts, voltage, electric field, etc.), and variability of data (constant for the whole run or changing with each event). The CERN ROOT TTrees and TFiles classes were chosen as the base for the data format, mainly due to their fast sequential and random i/o access, the convenience of use with ``Event-structured'' data, the responsiveness of the developers and likeliness of future existence and support. The described data format is already used for the storage and processing of the experimental data from the prototypes and simulated data coming from CoREAS \cite{Huege:2013vt} and ZHAireS \cite{ALVAREZMUNIZ2012325} simulators. The development of data processing and analysis tools has been significantly simplified with our Data-Oriented Interface that grants standard Pythonic access to our TTrees, which is not provided by the ROOT framework. Finally, the Analysis-Oriented Interface was designed to simplify the late stages of data analysis.

\bibliographystyle{JHEP}
\bibliography{lwp_icrc_2023}{}

\end{document}